\documentstyle[preprint,aps]{revtex}
\begin{document}
\draft
\title{Towards a Linear-Scaling DFT Technique: \\
The Density
Matrix Approach}
\author{E. Hern\'{a}ndez and M. J. Gillan}
\address{Physics Department, Keele University, Keele, Staffordshire
ST5 5BG, U.K.}
\author{C. M. Goringe}
\address{Materials Department, Oxford University, Oxford OX1 3PH, U.K.}
\date{\today}
\maketitle

\begin{abstract}
A recently proposed linear-scaling scheme for density-functional
pseudopotential calculations is described in detail. The method is based
on a formulation of density functional theory in which the ground state
energy is determined by minimization with respect to the density matrix,
subject to the condition that the eigenvalues of the latter lie in the
range [0,1]. Linear-scaling behavior is achieved by requiring that
the density matrix should vanish when the separation of its arguments
exceeds a chosen cutoff. The limitation on the eigenvalue
range is imposed by the method of Li, Nunes and Vanderbilt.
The scheme is implemented by calculating all terms in the energy
on a uniform real-space grid, and minimization is performed
using the conjugate-gradient method. Tests on a 512-atom Si system show that
the total energy converges rapidly as the range of the density
matrix is increased. A discussion of the relation between the present
method and other linear-scaling methods is given, and some
problems that still require solution are indicated.
\end{abstract}


\section{Introduction}

During the last decade, first-principles total-energy
methods based on density functional theory~(DFT) combined
with the pseudopotential method, have become
established as a major tool in the study of condensed
matter~\cite{dft_pseudo}.
The DFT-pseudopotential approach is now widely used for both static
and dynamic simulations on an enormous range of condensed-matter problems.
However, these methods suffer from a severe drawback in that their
computational cost generally increases as the
cube of the number of atoms in the system. This unfavorable
scaling limits the size of systems that can
be studied with current methods and today's computers to a few
hundred atoms at most. This $O(N^3)$ scaling appears in spite of
the fact that the complexity of the problem increases only
linearly with the system size. This observation suggests
that the unfavorable scaling of current methods is
a consequence of the way in which the electronic structure problem is
being addressed. Conventional methods rely either on diagonalization
of the Hamiltonian or orthonormalization of a set of occupied
orbitals, both of which are intrinsically $O(N^3)$ operations.
It is clear that more efficient methods in which the effort is proportional
to the number of atoms must be possible, and in recent years a
considerable effort has been devoted to finding
such `linear-scaling'
schemes~\cite{yang,baroni:giannozzi,galli:parrinello,mauri:galli:car,mauri:galli,ordejon1,ordejon2,kim:mauri:galli,li:nunes:vanderbilt,qiu,hierse:stechel,hernandez:gillan}.

The earliest linear-scaling scheme appears to be the `divide and
conquer' method of Yang~\cite{yang}. This obtains the electronic
density and hence the total energy by dividing the
system into overlapping sub-systems that can be treated independently.
The density is calculated for each sub-system
with conventional LCAO-DFT. The Hamiltonian for each sub-system, which
includes the potential due to the other sub-systems, is
diagonalized independently, thus avoiding the need to diagonalize
the full Hamiltonian.
This procedure is repeated
until self-consistency is achieved.
Baroni and Giannozzi~\cite{baroni:giannozzi} also proposed a scheme
that directly determines the electron density.
They do this by
discretizing the Kohn-Sham Hamiltonian on a real-space grid, and then
using the recursion method of Haydock, Heine and
Kelly~\cite{haydock:heine:kelly} to obtain the diagonal elements of
the Green's function, from which the electron density can be computed
by contour integration. In this case linear scaling results from the
fact that the continued fraction used to evaluate a particular
diagonal element of the Green's function can be truncated once a certain
neighborhood of each point has
been explored. This neighborhood is independent of the system size
for sufficiently large systems.

More recently, several new schemes that
resemble traditional first-principles methods have been
reported. Galli and Parrinello~\cite{galli:parrinello} pointed out
that some improvement could be achieved in the scaling of a
conventional DFT calculation by requiring spatial localization
of the electronic orbitals.
This localization was achieved by adding certain non-local constraining
terms to the Hamiltonian, or by using a filtering procedure. The total
energy can then be obtained as a functional of the localized orbitals
$| \phi_i \rangle$ and their {\em conjugate\/} orbitals
$\bar{| \phi_i \rangle} = \sum_j S^{-1}_{ji} | \phi_j \rangle$,
but in order to
obtain these conjugate orbitals, the overlap matrix $S$ has to be
inverted. Since spatial localization implies sparsity of $S$,
this can be achieved in $O(N^2)$ operations, so that
some improvement with respect to $O(N^3)$ is obtained. A step
further in this direction was made independently by Mauri, Galli and
Car~\cite{mauri:galli:car,mauri:galli} (hereafter referred to as MGC)
and by Ordej\'{o}n {\em et
al.\/}~\cite{ordejon1,ordejon2}. They introduced a new functional of
the occupied orbitals that possesses the same
ground state as the conventional energy functional, but with the added
advantage of leading naturally to orthogonal orbitals when minimized.
If this new functional is minimized with respect to orbitals that are
constrained to remain localized in chosen regions of space, as
suggested by Galli and Parrinello~\cite{galli:parrinello}, a linear
scaling method results. In the original formulation, the number of
orbitals entering the new functional is
equal to half the number of
electrons in the system. This restriction seems to lead to very slow
convergence, and to the appearance of spurious local minima in the
functional. This problem has been recently overcome by Kim, Mauri and
Galli~\cite{kim:mauri:galli}, by generalizing the functional so that
it depends on an arbitrary number of orbitals.

The linear-scaling scheme most relevant to the present work
is that put forward
by Li, Nunes and
Vanderbilt~\cite{li:nunes:vanderbilt} (hereafter referred
to as LNV) in the context
of tight-binding semi-empirical calculations. In this
method, linear scaling is achieved by taking advantage of the real
space localization properties of the density matrix, $\rho({\bf r},
{\bf r}')$. By introducing a spatial cutoff $R_c$ in $\rho$, such that
$\rho({\bf r}, {\bf r}')$ is set to zero if
$| {\bf r} - {\bf r}' | \geq R_c$,
the number of non-zero elements in $\rho$ increases only linearly with
the system size. The electronic structure problem is then formulated
as a minimization of the total energy with respect to the truncated
density matrix, subject to the constraints of idempotency ($\rho^2 =
\rho$) and correct trace ($2 \, \mbox{Tr} \, \rho = N_e$,
where $N_e$ is the number of electrons). The scheme of LNV
consists of an algorithm for imposing these
constraints that at the same time fulfils the goal of linear scaling.
The idempotency of $\rho$ is the most difficult constraint to impose,
and this scheme achieves it by expressing $\rho$ in terms of an
auxiliary matrix, which we denote in this paper by
$\sigma$. This is subjected to a {\em purifying\/}
transformation due to McWeeny~\cite{mcweeny}. If $\sigma$ is a
near-idempotent matrix, i.e. if its eigenvalues lie close to~0 or~1,
this transformation will return $\rho$ as a more nearly idempotent
matrix, and thus it is possible to minimize the total energy with
respect to $\sigma$ while ensuring the near idempotency of $\rho$.
By construction, the method is
variational (i.e. $\min E(R_c) \geq \min E(\infty)$), and it has been
shown that the convergence of calculated properties with the parameter
$R_c$ is fairly rapid~\cite{li:nunes:vanderbilt,qiu}. It is now being
widely used in tight-binding simulations of large systems.

Recently, the idea of working with the density matrix has been applied
to DFT linear scaling schemes. This has been done
independently by Hierse and Stechel~\cite{hierse:stechel} and by
Hern\'{a}ndez and Gillan~\cite{hernandez:gillan}. In both cases, the
density matrix is represented in real space as:
\begin{eqnarray}
\rho({\bf r},{\bf r}') = \sum_{\alpha \beta} \phi_\alpha ({\bf r}) \,
K_{\alpha \beta} \, \phi_\beta^\ast ({\bf r}') \; ,
\end{eqnarray}
where the $\phi_\alpha$ are a set of localized functions, and
$K_{\alpha \beta}$
is a symmetric matrix. The total energy is expressed in terms
of $\rho ( {\bf r}, {\bf r}^\prime )$, and
minimization is
carried out with respect to both the $\phi_\alpha$ and
the $K_{\alpha \beta}$. Hierse and Stechel~\cite{hierse:stechel}
use a number of functions $\phi_\alpha$ equal to the number of
occupied orbitals, but this restriction is not present in our
scheme. The consequences of this and other differences between the
two methods will be addressed later in this paper.

Previously, only a brief description of our method has
been published~\cite{hernandez:gillan}.
In this paper we give a detailed description of
the method, together with some illustrations of its practical
performance and a discussion of its relation to other methods.
In section 2, the method is outlined and its theoretical
foundations are discussed. The practical implementation of the method
is then described in section 3. The tests we have performed
to probe the practical usefulness of the scheme are presented in
section 4. In section 5, we assess what has been achieved and
we discuss possible future developments,
with particular attention to the problems that need to be overcome
before the method can be generally applied. Some of the
mathematical analysis is reported in an Appendix.

\section{Formulation of DFT in terms of the density matrix}
\label{sec:formulation}
\subsection{Density functional theory}
We need to recall briefly the principles of DFT~\cite{ksh}.
The total energy
$E_{\rm tot}$ of the system of valence electrons and atomic cores is
expressed as:
\begin{equation}
E_{\rm tot} = E_{\rm K} + E_{\rm ps} + E_{\rm H} + E_{\rm xc} +
E_{\rm M} \; ,
\label{eq:totalenergy}
\end{equation}
where the terms on the right are the kinetic, pseudopotential,
Hartree and exchange-correlation energies of the electrons, and
$E_{\rm M}$ is the Madelung energy of the cores. The first two energies
are:
\begin{eqnarray}
E_{\rm K} & = & 2 \sum_{i = 1}^{N} \langle \psi_i |
- \frac{\hbar^2}{2 m} \nabla^2 | \psi_i \rangle \nonumber \\
E_{\rm ps} & = & 2 \sum_{i = 1}^{N} \langle \psi_i |
{\hat{V}}_{\rm ps} | \psi_i \rangle \; ,
\end{eqnarray}
where $\psi_i$ are the Kohn-Sham (KS) orbitals, ${\hat{V}}_{\rm ps}$
is the total pseudopotential operator,
and $N = \frac{1}{2} N_e$ is the number of occupied orbitals.
The energies $E_{\rm H}$ and
$E_{\rm xc}$ can be written in terms of the electron number
density $n ( {\bf r} )$:
\begin{eqnarray}
E_{\rm H} & = & \frac{1}{2} e^2 \int d {\bf r} d {\bf r}^{\prime} \,
n ( {\bf r} ) n ( {\bf r}^{\prime} ) /
| {\bf r} - {\bf r}^{\prime} | \nonumber \\
E_{\rm xc} & = & \int d {\bf r} \, n ( {\bf r} ) \epsilon_{\rm xc}
\left( n ( {\bf r} ) \right ) \; ,
\end{eqnarray}
where for simplicity we assume the local density approximation
(LDA) for $E_{\rm xc}$, with $\epsilon_{\rm xc}$ the exchange-correlation
energy per electron.
The number density is:
\begin{equation}
n ( {\bf r} ) = 2 \sum_{i = 1}^{N} | \psi_i ( {\bf r} ) |^2 \; .
\end{equation}
The important principle for present purposes is that the true
ground-state energy and electron density are obtained by minimizing
$E_{\rm tot}$ with respect to the KS orbitals, subject to the
constraint that the latter are kept orthonormal.

In the standard formulation of DFT, which we have just summarized,
all the occupied orbitals are fully occupied. However, it is frequently
convenient, for physical, computational or formal reasons, to
generalize the theory so that orbitals can be partially occupied.
Spatial orbital $\psi_i ( {\bf r} )$, rather than containing
2 electrons, may now contain $2 f_i$ electrons, where
the occupation number $f_i$ lies in the range $0 \leq f_i \leq 1$.
The number density $n ( {\bf r} )$
now becomes:
\begin{equation}
n ( {\bf r} ) = 2 \sum_{i} f_i | \psi_i ( {\bf r} ) |^2 \; ,
\end{equation}
and the kinetic and pseudopotential energies are:
\begin{eqnarray}
E_{\rm K} & = & 2 \sum_i f_i \langle \psi_i |
- \frac{\hbar^2}{2 m} \nabla^2 | \psi_i \rangle \nonumber \\
E_{\rm ps} & = & 2 \sum_i f_i \langle \psi_i |
{\hat{V}}_{\rm ps} | \psi_i \rangle \; .
\end{eqnarray}
The expressions for $E_{\rm H}$ and $E_{\rm xc}$ in terms of
$n ( {\bf r} )$ are unchanged.

The usual physical reason for making this generalization is
that one wishes to treat the electrons at a non-zero temperature,
in which case the $f_i$ are Fermi-Dirac occupation numbers~\cite{mermin};
computationally, the generalization is sometimes made in order to
get rid of the troublesome discontinuity at the Fermi level
in metallic systems~\cite{gillan:89,grumbach:94}.
Our reason for considering it here is that
it will be relevant to the density matrix formulation.
We shall assume that if $E_{\rm tot}$ is minimized both
with respect to the $\psi_i$ (subject to orthonormality) and
with respect to the $f_i$ (subject to the restriction
$0 \leq f_i \leq 1$ and the condition that the sum $f_i$
be equal to $\frac{1}{2} N_e$),
then we arrive at exactly the ground state
that is obtained by the more usual
minimization with respect to fully occupied states $\psi_i$.
Another way of putting this is that the energy cannot be reduced
below the normal ground state by allowing partial
occupation.

Now we turn to the density matrix, which is defined by
\begin{equation}
\rho ( {\bf r}, {\bf r}^{\prime} ) =
\sum_i f_i \psi_i ( {\bf r} ) \psi_i^{*} ( {\bf r}^{\prime} ) \; .
\label{eq:rho_def}
\end{equation}
It follows from this definition that $\rho ( {\bf r},  {\bf r}^{\prime} )$
is a Hermitian operator whose eigenvalues are all in the interval
$[ 0, 1 ]$. The converse is also true: a Hermitian operator
$\rho ( {\bf r} , {\bf r}^{\prime} )$ whose eigenvalues are $f_i$ and
whose eigenfunctions are $\psi_i ( {\bf r} )$
can be written as in
equation (\ref{eq:rho_def}).
In terms of
such an operator $\rho ( {\bf r} , {\bf r}^{\prime} )$,
let the kinetic energy, pseudopotential energy and number density
be defined as:
\begin{eqnarray}
E_{\rm K} & = & - \frac{\hbar^2}{m} \int d {\bf r} \;
\left( \nabla_r^2
\rho ( {\bf r}, {\bf r}^{\prime} ) \right)_{{\bf r} = {\bf r}^{\prime}}
\nonumber \\
E_{\rm ps} & = & 2 \int d {\bf r} d {\bf r}^{\prime} \;
V_{\rm ps} ( {\bf r}^{\prime}, {\bf r} )
\rho ( {\bf r}, {\bf r}^{\prime} ) \nonumber \\
n ( {\bf r} ) & = & 2 \rho ( {\bf r}, {\bf r} ) \; ,
\label{eq:traces}
\end{eqnarray}
with $E_{\rm H}$ and $E_{\rm xc}$ expressed in the usual way in terms
of $n ( {\bf r} )$. It follows from what we have said before
that if $E_{\rm tot}$ is minimized with respect to
$\rho( {\bf r}, {\bf r}^{\prime} )$ subject to the condition that
the eigenvalues of the latter are in the required interval
and add up to $\frac{1}{2} N_e$, then we arrive at
the usual ground state. This
is the density matrix formulation of DFT.

\subsection{Localization of the density matrix}
Since DFT is variational, any restriction placed on the class of
density matrices $\rho ( {\bf r}, {\bf r}^{\prime} )$ that can
be searched over has the effect of raising the minimum energy
$E_{\rm min}$ above its true ground-state value $E_0$; progressive
relaxation of such a restriction makes $E_{\rm min}$ tend to $E_0$.
Now in general the density matrix in the true ground state tends to
zero as the separation of its arguments $| {\bf r} -
{\bf r}^{\prime} |$ increases. This strongly suggests the usefulness
of estimating $E_0$ by searching over $\rho ( {\bf r}, {\bf r}^{\prime} )$
with the restriction that:
\begin{equation}
\rho ( {\bf r}, {\bf r}^{\prime} ) = 0 \; , \; | {\bf r} -
{\bf r}^{\prime} | > R_c \; ,
\label{eq:dm_cutoff}
\end{equation}
where $R_c$ is a chosen cutoff radius. The resulting estimate
$E_{\rm min} ( R_c )$ will tend to $E_0$ from above as
$R_c \rightarrow \infty$. The manner in which $\rho ( {\bf r},
{\bf r}^{\prime} )$ goes to zero at large separations depends
on the electronic structure of the system, and particularly
on whether there is a gap between the highest occupied and
lowest unoccupied states. It is rigourously established that
in one-dimensional systems having a gap $\rho$ decays
exponentially with separation, while in gapless systems it
decays only as an inverse power~\cite{kohn:59}.
It is presumed that three-dimensional
systems behave similarly. This suggests -- though to our knowledge
it is unproven -- that $E_{\rm min} ( R_c ) \rightarrow E_0$
exponentially for insulators and algebraically for metals.

Clearly in practical calculations we cannot work directly with
a six-dimensional function $\rho ( {\bf r}, {\bf r}^{\prime} )$,
even if it vanishes beyond a chosen radius. It is essential that
$\rho$ be separable, i.e. representable in the form:
\begin{equation}
\rho ( {\bf r}, {\bf r}^{\prime} ) =
\sum_{\alpha \beta} \phi_{\alpha} ( {\bf r} ) K_{\alpha \beta}
\phi_{\beta} ( {\bf r}^{\prime} ) \; .
\label{eq:separable}
\end{equation}
For practical purposes, there must be only a finite number of
$\phi_{\alpha} ( {\bf r} )$ functions, which will be referred
to as support functions. For $\rho$ to be Hermitian, we must
require that the matrix $K_{\alpha \beta}$ be Hermitian.
The restriction to a finite number of support functions is
equivalent to the condition that $\rho$ have only this number
of non-zero eigenvalues, and this is the essence of the separability
requirement. With this, we now have two independent restrictions
on $\rho$: localization and separability.
The localization of $\rho$ can be imposed by requiring that
the support functions be non-zero only within chosen regions, which we
call the support regions, and that the coefficients $K_{\alpha \beta}$
vanish if the separation of the support regions of $\phi_\alpha$
and $\phi_\beta$ exceeds a chosen cutoff.

We now have a general framework for linear-scaling DFT schemes.
In practical calculations, the $\phi_\alpha$ functions will be
represented either as a linear combination of basis functions,
or simply by numerical values on a grid. Either way, the amount of
information contained in a support function will be independent
of the size of the system.
The amount of information in the support
functions will then scale linearly with the size of the system,
and the number of $K_{\alpha \beta}$ coefficients will scale in
the same way. This in turn implies that the electron density
$n( {\bf r} )$ and all the terms in the total energy can be calculated
in a number of operations which scales linearly with system
size.

\subsection{Eigenvalue range of the density matrix}
In this general scheme, the ground state is determined by searching
over support functions and $K_{\alpha \beta}$ matrices. However,
it is essential that this search be confined to those $\phi_\alpha$
and $K_{\alpha \beta}$ for which the eigenvalues of
$\rho ( {\bf r}, {\bf r}^\prime )$ lie in the interval [0,1].
This is a troublesome condition to impose, because we certainly
do not wish to work directly with these eigenvalues. We can achieve
what we want by expressing $\rho$ in a form that satisfies the
condition automatically.

The scheme developed in this paper is the DFT analogue of the tight-binding
scheme of LNV~\cite{li:nunes:vanderbilt}.
We write the density matrix as:
\begin{equation}
\rho = 3 \sigma * \sigma - 2 \sigma * \sigma * \sigma \; ,
\end{equation}
where $\sigma ( {\bf r}, {\bf r}^\prime )$ is an auxiliary function.
(The asterix here indicates the continuum analogue of matrix
multiplication. For arbitrary two-point functions
$A ( {\bf r}, {\bf r}^\prime )$ and $B ( {\bf r}, {\bf r}^\prime )$,
we use the notation $C = A * B$ as a short-hand for the statement:
\begin{equation}
C( {\bf r}, {\bf r}^\prime ) =
\int d {\bf r}^{\prime \prime} \, A ( {\bf r}, {\bf r}^{\prime \prime} )
B ( {\bf r}^{\prime \prime} , {\bf r}^{\prime} ) \; .)
\end{equation}
The reason this works is that the eigenvalues $\lambda_\rho$ of
$\rho$ automatically satisfy $0 \leq \lambda_\rho \leq 1$
provided the eigenvalues $\lambda_\sigma$ of $\sigma$ are in the
range $- \frac{1}{2} \leq \lambda_\sigma \leq \frac{3}{2}$; in addition,
$\lambda_\rho$ has turning points at the values 0 and 1.
Since the ground state is obtained when $\lambda_\rho = 0$ or 1, there
is a natural mechanism whereby variation of $\sigma$ drives $\rho$ towards
idempotency.

To obtain the separable form of $\rho$ (Eq.~(\ref{eq:separable})), we write:
\begin{equation}
\sigma ( {\bf r}, {\bf r}^\prime ) = \sum_{\alpha \beta}
\phi_\alpha ( {\bf r} ) L_{\alpha \beta} \phi_\beta ( {\bf r}^\prime ) \; ,
\label{eq:sigma_def}
\end{equation}
which implies the matrix relation:
\begin{equation}
K = 3LSL - 2LSLSL \; ,
\label{eq:k_of_l}
\end{equation}
where $S_{\alpha \beta}$ is the overlap matrix of support functions:
\begin{equation}
S_{\alpha \beta} = \int d {\bf r} \, \phi_\alpha ( {\bf r} )
\phi_{\beta} ( {\bf r} ) \; .
\end{equation}
The ground state is now obtained by minimizing $E_{\rm tot}$
with respect to the $\phi_\alpha$ and the $L_{\alpha \beta}$
matrix, with the $K_{\alpha \beta}$ matrix given by Eq.~(\ref{eq:k_of_l}).
In the practical calculations reported later, the $\phi_\alpha$ are
non-zero only inside spherical regions of radius $R_{\rm reg}$, and
the $L_{\alpha \beta}$ are non-zero only if the centres of the regions
$\alpha$ and $\beta$ are separated by less than a cutoff distance $R_{\rm L}$.

It will be useful for the purposes of later discussion to note how
a closely related scheme leads back to the
MGC method~\cite{mauri:galli:car}. This
scheme is obtained by writing:
\begin{equation}
\rho = \sigma * ( 2 - \sigma ) \; ,
\end{equation}
where $\sigma$ is required to be positive semi-definite. Since the
eigenvalues $\lambda_\sigma$ can be expressed as $\kappa_\sigma^2$
where $\kappa_\sigma$ is real, the eigenvalues of $\rho$ are given
by:
\begin{equation}
\lambda_\rho = \lambda_\sigma ( 2 - \lambda_\sigma ) =
\kappa_\sigma^2 ( 2 - \kappa_\sigma^2 ) \; .
\end{equation}
This quartic function lies in the range [0,1] for $| \kappa_\sigma |
\leq 2^{1/2}$ and has turning points when $\lambda_\rho = 0$ and 1. This
give an alternative mechanism for driving $\rho$ towards idempotency.
With $\sigma$ given, as before, by Eq.~(\ref{eq:sigma_def}), it is
straightforward to show that $\sigma$ is positive semi-definite
if and only if the matrix $L_{\alpha \beta}$ is positive
semi-definite, and this is equivalent to the condition that
$L_{\alpha \beta}$ be expressible as:
\begin{equation}
L_{\alpha \beta} = \sum_s b_\alpha^{(s)} b_\beta^{(s)} \; .
\end{equation}
The result is that $\sigma ( {\bf r}, {\bf r}^\prime )$ must have
the form:
\begin{equation}
\sigma ( {\bf r}, {\bf r}^\prime ) = \sum_s \chi^{(s)} ( {\bf r} )
\chi^{(s)} ( {\bf r}^\prime ) \; ,
\end{equation}
where:
\begin{equation}
\chi^{(s)} ( {\bf r} ) = \sum_{\alpha} b_\alpha^{(s)}
\phi_\alpha ( {\bf r} ) \; .
\end{equation}
Following arguments presented by Nunes and Vanderbilt~\cite{nunes:vanderbilt}
in the tight-binding
context, it can now be shown that this scheme is exactly
equivalent to the linear-scaling DFT scheme of MGC.

\section{Practical implementation of the method}
\label{sec:details}
\subsection{The real-space grid}
We now give a prescription for the calculation of the energy functional,
and of its derivatives with respect to the support
functions $\phi_\alpha$ and
the $L_{\alpha \beta}$ parameters, and we describe how minimization
of the energy
can be carried out in practice. Central to our implementation of the
method described in the previous section is the use of a regular
cubic real-space grid,
spanning the whole system under study.
There have been a number of recent implementations of conventional
DFT-pseudopotential calculations using real-space
grids~\cite{chelikowsky,briggs,gygi,zumbach}.

The support functions are represented by their values at the grid
points. Since these functions are required to be spatially
localized, they have non-zero values only on the grid points
inside the localization regions. In the present work, these
regions are chosen to be spherical, and their centres are at the atomic
positions. Real-space integration is replaced by summation over grid
points, so that e.g. the overlap matrix elements are calculated as:
\begin{eqnarray}
S_{\alpha \beta} \simeq \delta \omega \sum_{{\bf r}_\ell} \phi_\alpha
({\bf r}_\ell) \, \phi_\beta({\bf r}_\ell) \; ,
\label{eq:overlap}
\end{eqnarray}
where the sum goes over the set of grid points ${\bf r}_\ell$
common to the localization regions of both $\phi_\alpha$
and $\phi_\beta$, and $\delta \omega$ is the volume per grid point.

The action of the kinetic energy operator on the support
functions is evaluated using a finite difference technique.
To $n$th order in the grid spacing, $h$, we have that
\begin{eqnarray}
\frac{\partial^2 \phi_\alpha}{\partial x^2} (n_x, n_y, n_z) \simeq
\frac{1}{h^2} \sum_{m=-n}^{n} C_{\mid m \mid}
\phi_\alpha (n_x+m,n_y,n_z),
\label{eq:finitedifference}
\end{eqnarray}
where $n_x, n_y$ and $n_z$ are integer indices
labelling grid point ${\bf r}_\ell$,
and the coefficients $C_{\mid m \mid}$ can be calculated beforehand.
Equivalent expressions can be used for $\partial^2 \phi_\alpha/\partial y^2$
and $\partial^2 \phi_\alpha /\partial z^2$, and it is thus possible to
evaluate $\nabla^2 \phi_\alpha$ approximately at each grid point.
{}From Eqs.~(\ref{eq:traces}) and (\ref{eq:separable}),
the kinetic energy is given by:
\begin{equation}
E_{\rm K} = 2 \sum_{\alpha \beta} K_{\alpha \beta} T_{\beta \alpha} \; ,
\end{equation}
where
\begin{equation}
T_{\beta \alpha} = - \frac{\hbar^2}{2 m} \int d {\bf r} \,
\phi_\beta ( {\bf r} ) \nabla_r^2 \phi_\alpha ( {\bf r} ) \; .
\end{equation}
Once $\nabla^2 \phi_\alpha ({\bf r})$ has been evaluated at each grid point
using Eq.~(\ref{eq:finitedifference}), the $T_{\alpha \beta}$
matrix elements are calculated by summing over grid points, just
as for $S_{\alpha \beta}$ (see Eq.~(\ref{eq:overlap})).

In order to evaluate the exchange-correlation, Hartree and pseudopotential
contributions to the total energy, we first need to evaluate the electron
density at each grid point. From Eqs.~(\ref{eq:traces}) and
(\ref{eq:separable}),
the density at grid
point ${\bf r}_\ell$ is:
\begin{eqnarray}
n({\bf r}_\ell) = 2 \sum_{\alpha \beta} \phi_\alpha({\bf r}_\ell)
K_{\alpha \beta}
\phi_\beta({\bf r}_\ell) \; .
\label{eq:density}
\end{eqnarray}
{}From this, it is straightforward to evaluate the exchange-correlation
energy by summing the quantity $n({\bf r}_\ell)
\epsilon_{xc}[n({\bf r}_\ell)]$
over grid points.
The exchange-correlation potential
$\mu_{xc}$ can also be calculated at each point, and is given
as
\begin{eqnarray}
\mu_{xc}({\bf r}_\ell) = \frac{d}{dn} \left\{
n({\bf r}_\ell) \epsilon_{xc}[n({\bf r}_\ell)] \right\} \; .
\end{eqnarray}

To obtain the Hartree energy and potential we use
the fast Fourier transform (FFT) method
to transform the calculated electronic density into reciprocal space, thus
obtaining its Fourier components $\hat{n}_{\bf G}$. The Hartree energy is
then given as
\begin{eqnarray}
E_{\rm H} = 2 \pi \Omega e^2
\sum_{{\bf G} \neq {\bf 0}} | \hat{n}_{\bf G} |^2 / G^2 \; ,
\end{eqnarray}
where $\Omega$ is the volume of the simulation cell.
The Hartree potential in reciprocal space is:
\begin{eqnarray}
\hat{V}_H({\bf G}) = 4 \pi \Omega e^2
\hat{n}_{\bf G} / G^2 \; .
\end{eqnarray}
This can be constructed on the reciprocal-space grid, and transformed
to obtain the Hartree potential in real space. FFT
is, of course, an $O(N \log_2 N )$ operation rather than an $O(N)$
operation, but the difference is negligible for present purposes.

We restrict ourselves here to local pseudopotentials,
so that the value of the total pseudopotential $V_{\rm ps} ( {\bf r}_\ell )$
at grid point ${\bf r}_\ell$ is formally given by:
\begin{equation}
V_{\rm ps} ( {\bf r}_\ell ) = \sum_I v_{\rm ps} ( | {\bf r}_\ell -
{\bf R}_I | ) \; ,
\label{eq:pseudo_sum}
\end{equation}
where $v_{\rm ps} (r)$ is the ionic pseudopotential and ${\bf R}_I$
is the position of ion $I$. In practice, however,
$V_{\rm ps} ( {\bf r}_\ell )$
cannot be calculated like this, because $v_{\rm ps} (r)$ has a
Coulomb tail $-Z |e|^2 /r$ at large $r$, where $Z$ is the core charge.
In order to obtain a linear-scaling algorithm for $E_{\rm ps}$,
we proceed as follows. The ionic pseudopotential is represented
as the sum of the Coulomb potential due to a Gaussian charge
distribution $\eta (r)$ and a short-range potential $v_{\rm ps}^0 (r)$.
The total charge in $\eta (r)$ is $Z |e|$, and the distribution
is given by:
\begin{equation}
\eta (r) = \ Z |e| ( \alpha / \pi )^{3/2} \exp ( - \alpha r^2 ) \; ,
\end{equation}
where the parameter $\alpha$ governs the rate of decay of the
Gaussian. We therefore have:
\begin{equation}
v_{\rm ps} (r) = - \frac{Z |e|^2}{r} \, {\rm erf} ( \alpha^{1/2} r ) +
v_{\rm ps}^0 (r) \; .
\end{equation}
The part of $V_{\rm ps}$ coming from $v_{\rm ps}^0$ can now be
calculated as a direct sum over ions, as in
Eq.~(\ref{eq:pseudo_sum}). Since
$v_{\rm ps}^0$ can be neglected beyond a certain radius, this part
of the calculation scales linearly. The part of $V_{\rm ps}$ coming from
the array of Gaussians can be treated in exactly the same way as
the Hartree potential. The pseudopotential energy is then calculated
by summation over the real-space grid:
\begin{equation}
E_{\rm ps} = \delta \omega \sum_{\ell} V_{\rm ps} ( {\bf r}_{\ell} )
n ( {\bf r}_{\ell} ) \; .
\end{equation}

\subsection{Derivatives and minimization}
Once the contributions to the total energy have been obtained
as outlined above, we need to vary
both $L_{\alpha \beta}$ and $\phi_\alpha$ in order to minimize it.
The $L_{\alpha \beta}$ and $\phi_\alpha$ are independent variables,
and the problem breaks naturally into two separate minimizations that
can be carried out in an alternating manner:
one with respect to $L_{\alpha \beta}$ with fixed $\phi_\alpha$,
and the other with respect to $\phi_\alpha$ with fixed $L_{\alpha \beta}$.
Indeed, the choice of object function can be different
for the two types of variation, and when minimizing with respect to
the $L_{\alpha \beta}$ we find it more convenient to take
$\Omega = E_{\rm tot} - \mu N_e$ as our object function, where $\mu$ is the
chemical potential and $N_e$ is the electron
number. We return to this point below.

Expressions for the derivatives with respect to $L_{\alpha \beta}$
and $\phi_\alpha$ are obtained in Appendix A.
The partial derivative of $\Omega$ with respect to $L_{\alpha \beta}$
is given by
\begin{eqnarray}
\frac{\partial \Omega}{\partial L_{\alpha \beta}} =
\left[ 6(SLH' + H'LS) - 4 (SLSLH' + SLH'LS +H'LSLS)\right]_{\alpha \beta}
\label{eq:partialab}
\end{eqnarray}
where $H' = H - \mu S$, and $H$ is the matrix representation of the
KS Hamiltonian in the support function representation.
It is worth noting that this expression is exactly the same as
would be obtained in a non-orthogonal tight-binding
formalism~\cite{non_orthog_tb}. There is, however, one important
difference: in self-consistent~DFT calculations the Hamiltonian matrix
elements depend on $L_{\alpha \beta}$ through the electronic density
$n({\bf r})$.
The partial derivative of the total energy with respect to $\phi_\alpha$
at grid point ${\bf r}_\ell$ is given by
\begin{eqnarray}
\frac{\partial E_{\rm tot}}{\partial \phi_\alpha ({\bf r}_\ell)} =
4 \delta \omega
\sum_\beta \left[ K_{\alpha \beta} \hat{H} +
3 (LHL)_{\alpha \beta} - 2(LSLHL + LHLSL)_{\alpha \beta}\right]
\phi_\beta({\bf r}_\ell),
\label{eq:partialphi}
\end{eqnarray}
where $\hat{H}$ is the Kohn-Sham operator, which is made to act on
support function $\phi_\beta$.

It is important to notice that because of the spatial localization of
the support functions, and the finite range of $L$,
all the matrices involved in the calculation of these
derivatives are sparse, when the system is large enough.
Provided this sparsity is exploited in the computational
scheme, the method scales linearly with the size of the system.

In the scheme of LNV~\cite{li:nunes:vanderbilt},
it is proposed to work
at constant chemical potential, rather than at constant electron number.
We prefer to maintain the electron number constant.
The variations with respect to $L_{\alpha \beta}$ and
$\phi_\alpha$ will in general cause the electron number to differ from the
correct value, and it is therefore necessary to correct this effect
as the minimization proceeds. We achieve this in the following manner:
during the minimization with respect to $L$, the current search direction
is projected so that it is tangential
to the local surface of constant $N_e$,
i.e. perpendicular to $\nabla_L N_e$ at the current position. This
ensures that the minimization along this direction will cause only a
small change in $N_e$, and it is expected that at the new minimum
$N_e$ will differ only slightly from the required value.
In any case, it is possible to
return to a position as close as desired to the
constant $N_e$ surface by following
the local gradient $\nabla_L N_e$. If the value of the chemical potential
$\mu$
is appropriately chosen, this correction step can be carried out without
losing the reduction in $\Omega$
obtained by performing the line minimization, and
this is why we prefer to take $\Omega$ as the object function instead of the
total energy, when minimizing with respect to $L$. We find that this
scheme is capable of maintaining the electron number close to its correct
value throughout the minimization, and is also simple to implement. The
gradient $\nabla_L N_e$ has elements
\begin{eqnarray}
\frac{\partial N_e}{\partial L_{\alpha \beta}} =
12 (SLS - SLSLS)_{\alpha \beta} \; ,
\end{eqnarray}
which, as all other gradients discussed earlier, can be calculated in
$O(N)$ operations. Minimization with respect to
$\phi_\alpha({\bf r}_\ell)$ will also have
the effect of changing the electron
number. However, given that the two types of variation are performed
alternately, the correction during the $L$ minimization is
sufficient to counteract this effect.

Given that variation of $L_{\alpha \beta}$ causes the electronic
density to change, and this in turn implies that the Hamiltonian
matrix elements change, it would seem necessary to update
the Hamiltonian at each step of the minimization with respect to $L$.
However, we find that this can be avoided by considering $H$ fixed
during this part of the minimization. Strictly speaking, if $H$ is
held fixed while $L$ is varied, we are not minimizing $\Omega =
E_{\rm tot} - \mu N$ but rather $\Omega' = E' - \mu N$, where $E'$ is
given by
\begin{eqnarray}
E' = \mbox{Tr} [( 6 LSL - 4 LSLSL) H] \; .
\end{eqnarray}
If this minimization were carried out through to convergence, this would
be equivalent to diagonalizing $H$ in the representation of the
current support functions. At convergence, it will be found in general
that $L$ and $H$ are not mutually consistent, and if consistency is
required, one needs to update $H$ and repeat the minimization,
iterating this cycle until consistency was achieved. This is not
necessary in practice, because $H$ will be updated at the next
variation with respect to the support functions.
The minimization of $\Omega'$ has practical advantages in that it
avoids the updating of the Hamiltonian at each step, and, because of
its construction, it is a cubic polynomial in every possible search
direction, so it is possible to find the exact location of line minima
during its minimization.

The minimization with respect to $\phi_\alpha({\bf r}_\ell)$ can be carried
out by simply moving along the gradient
$\partial E_{\rm tot} / \partial \phi_\alpha({\bf r}_\ell)$
Eq.~(\ref{eq:partialphi}) (steepest descents)
or by using this expression to construct mutually conjugate directions
(conjugate gradients).

\section{Test calculations}
In order to test our $O(N)$ DFT scheme, we have performed calculations
on a system of 512 Si atoms treated using a local pseudopotential.
The purpose of these tests is to find out how the total energy
depends on the two spatial cutoff radii: the support-region
radius $R_{\rm reg}$, and the $L$-matrix cutoff radius $R_{\rm L}$.
The practical usefulness of the scheme, and the size of system for
which linear-scaling behavior is attained depend on the rate of
convergence of $E_{\rm tot}$ to its exact value as $R_{\rm reg}$
and $R_{\rm L}$ are increased. Here, `exact' refers only to the
absence of errors due to the truncation of $\rho ( {\bf r},
{\bf r}^\prime )$; other sources of inexactness, such as
the use of a discrete grid and a local pseudopotential, are of no
concern here.

The system treated is a periodically repeating cell containing
512 atoms of diamond-structure Si having the experimental
lattice parameter (5.43 \AA). The local pseudopotential
is the one constructed by Appelbaum and Hamann~\cite{appelbaum}, which is
known to give a satisfactory representation of the self-consistent
band structure. The LDA exchange-correlation energy is calculated
using the Ceperley-Alder formula~\cite{ceperley}.
We use a grid spacing of 0.34 \AA,
which is similar to the spacing typically used in pseudopotential
plane-wave calculations on Si, and is sufficient to give
reasonable accuracy. The second derivatives of the $\phi_{\alpha}$
needed in the calculation of $E_{\rm K}$ are computed using the
second-order formula given in Eq.~(\ref{eq:finitedifference}).

A support region is centred on every atom, and each such region contains
four support functions. One can imagine that these support functions
correspond roughly to the single 3$s$ function and the three
3$p$ functions that would be used in a tight-binding
description, but we stress that nothing obliges us to work
with this number of support functions. In keeping with the tight-binding
picture, the initial guess for the support functions is taken to be
a Gaussian multiplied by a constant, $x$, $y$ or $z$, so
that the functions have the symmetry of $s$ and $p$ states. As an initial
guess for the $L$-matrix, we take the quantity $2 I - S$, where
$S$ is the overlap matrix calculated for the initial support
functions. This guess for $L$, which represents the expansion
of $S^{-1} \equiv ( I - ( I - S ))^{-1}$ to first order,
is crude, and does not yield the correct value of ${\rm Tr} \, \rho$.
This error is corrected by displacing $L$ iteratively along the
gradient $\nabla_L N_e$ until $N_e$ is within a required
tolerance of the correct value.

The initial guesses for the $\phi_\alpha$ and the $L_{\alpha \beta}$
define the initial Hamiltonian and overlap matrices. From
this starting point, we make a number of conjugate-gradient
line searches to minimize $\Omega$ by varying $L$, with the Hamiltonian
and overlap matrices held fixed. This is followed by a sequence of
line searches in which the $\phi_\alpha$ are varied. We refer to the
sequence of $L$ moves followed by a sequence of $\phi$ moves
as a {\em cycle}. The entire energy minimization consists
of a set of cycles. In practice, we have found that cycles
consisting of five $L$ moves and two $\phi$ moves
work satisfactorily, and that $E_{\rm tot}$ is converged to within
$10^{-4}$~eV/atom after typically 50-60 cycles. This would not be an
efficient rate of convergence for routine applications, but is more
than adequate for present purposes.

Our test calculations confirm our earlier finding~\cite{hernandez:gillan}
that for the Si perfect
crystal $E_{\rm tot}$ is already quite close to its exact value
when $R_{\rm L}$ = 5.0~\AA. We have therefore used
this value of $R_{\rm L}$ to make calculations of $E_{\rm tot}$ as
a function of $R_{\rm reg}$ (see fig. 1a). The results show that $E_{\rm tot}$
converges very quickly with increasing $R_{\rm reg}$, and that it is
within $\sim$~0.1~eV of its fully converged value for $R_{\rm reg}$ =
3.05~\AA. In order to show how $E_{\rm tot}$ depends on $R_{\rm L}$,
we present a series of results at the two region radii $R_{\rm reg}$ =
2.04 and 2.38~\AA\ (see fig. 1b). These results indicate that there is only a
slow variation with $R_{\rm L}$ and that this variation is almost the same
for different values of $R_{\rm reg}$. This means that it is possible to
converge the total energy to satisfactory accuracy with easily manageable
spatial cutoffs.

It is interesting to know the form of the support functions for the
self-consistent ground state. These are shown in fig.~2 for the case
$R_{\rm reg}$ = 3.05, $R_{\rm L}$ = 5~\AA. The support functions shown
here are the first (initially $s$~Gaussian) and second
($p_x$~Gaussian). Profiles of the support functions along the~[100],
[110] and~[111] directions are shown. The support functions are seen
to be symmetric with respect to the center of the support region ($r =
0$) along the [100] and [110] directions. Along the [111] direction
there is a slight asymmetry resulting from the presence of a nearest
neighbour ion, which lies at 2.35~\AA\ from the origin in the positive
direction. Remarkably, the $s$-like support function seems to be
almost perfectly spherically symmetric, except near the peak at
$r \approx 0.8$~\AA. It is encouraging to see that the support
functions go rather smoothly to zero at the region boundary, and this
confirms that the boundary is having little effect on the results.

\section{Discussion}
We have tried to do three things in this work: to develop the
basic formalism needed to underpin $O(N)$ DFT-pseudopotential
methods; to implement one such method and identify the main technical
issues in doing so; and to present the results of tests on a simple
but important system, which allow us to gauge the usefulness
of the method. We have shown that a rather general class of $O(N)$
DFT-pseudopotential methods can be based on a formulation of DFT in
terms of the density matrix, and that this formulation is equivalent
to commonly used versions of DFT that operate with fractional occupation
numbers. From this viewpoint, the key challenge is to ensure
that the eigenvalues of the variable density matrix lie between 0 and 1,
and we have seen that the method of
LNV~\cite{li:nunes:vanderbilt} gives a way of doing
this. The implementation of the basic ideas has been achieved by
performing all calculations in real space, with the DFT integrals
approximated by sums on a grid -- except for the use of FFT to treat
the Hartree term. An alternative here would be to work with
atomic-like basis functions, but we note that the use of a grid
preserves an important link with conventional plane-wave methods,
as will be analysed in more detail elsewhere. Our test results
on perfect-crystal Si show that the total energy converges rapidly as the
real-space cutoffs are increased, and that it is straightforward
to achieve a precision comparable with that of normal plane-wave calculations.

An important question for any $O(N)$ method is the system size at
which it starts to beat a standard $O(N^3 )$ method -- a plane-wave
method in the present case. This will clearly depend strongly on the system,
but even for Si it is too soon to answer it on the basis of practical
calculations. The cross-over point depends on the prefactor in the
linear scaling, and this is strongly affected by the efficiency of
the coding. All we have attempted to do here is to address the problem
of achieving $O(N)$ behavior. The question of the prefactor is a separate
matter, which will need separate investigation.

It should be clear that there is much more to do before the present
methods can be routinely applied to real problems. We have deliberately
not discussed in detail the problems of doing
calculations on a real-space grid.
Such problems have been discussed outside the linear-scaling context
in several recent papers~\cite{chelikowsky,briggs}, and
it should be possible to apply
the advances reported there to $O(N)$ DFT calculations. In particular,
curvilinear grids~\cite{gygi,zumbach}
for the treatment of strongly attractive pseudopotentials
are likely to be very important for $O(N)$ calculations. We have also not
discussed here the calculation of forces on the atoms, the problems
that may arise when the boundaries of support regions cross grid points,
and the general question of translational invariance
within grid-based techniques.

We have noted already that our method is related to other recently
proposed methods. As shown in sec. 2, the Mauri-Galli-Car scheme is
obtained from ours by taking an alternative polynomial expression
for the density matrix $\rho$ in terms of the auxiliary matrix $\sigma$.
It would clearly be interesting to repeat the calculations done here using
this approach. In a sense, this is what has already been done by Hierse
and Stechel~\cite{hierse:stechel}, except that instead of
performing calculations on a grid,
they use a minimal atomic basis set. The hydrocarbon systems used
by them for test purposes are also rather different from the Si
crystal we have studied.

Finally, we note that our linear-scaling scheme is intended for
calculations on very large systems, and this means that parallel
implementation will play a key role. The test calculations we have
presented were, in fact, performed on a massively parallel machine,
and the parallel-coding techniques we have developed will be described
in a separate paper.

\section*{Acknowledgments}
The work of CMG is supported by the High Performance Computing
Initiative (HPCI) under grant GR/K41649, and the work of EH by
EPSRC grant GR/J01967. The major calculations were done on the
Cray T3D at Edinburgh Parallel Computing
Centre using an allocation of time from the HPCI. Code
development and subsidiary analysis were made using local
hardware funded by EPSRC grant GR/J36266. We gratefully
acknowledge useful discussions with D. Vanderbilt.

\appendix

\section{Derivatives of the total energy}

\label{sec:derivatives}
We derive here expressions for the derivatives
$\partial E_{\rm tot} /\partial L_{\alpha \beta}$ and $\delta E_{\rm tot} /
\delta \phi_\alpha({\bf r})$.

\subsection{Derivative with respect to $L_{\alpha \beta}$}

In DFT, the total energy Eq.~(\ref{eq:totalenergy}) has two types of
contributions: those that can be written as the trace of some operator
acting on the density matrix, as is the case for the kinetic and
pseudopotential energies (see Eq.~(\ref{eq:traces})), and those
that depend only on the diagonal elements of
$\rho$, i.e. the electron density,
namely the Hartree and exchange-correlation energies. The
Madelung term, $E_M$, does not depend on either $L_{\alpha \beta}$ or
$\phi_{\alpha}$, so it will make no contribution to the variation in
total energy as these are changed. Denoting by $E_c$ the kinetic
or pseudopotential contribution to the energy, we have:
\begin{eqnarray}
E_c = 2 \sum_{\gamma \delta} \left[ 3(LSL)_{\gamma \delta} -
2(LSLSL)_{\gamma \delta} \right] C_{\delta \gamma} \; ,
\label{eq:energycont}
\end{eqnarray}
where
\begin{eqnarray}
C_{\delta \gamma} = \int d {\bf r} \; \phi_\delta ({\bf r})
\left( -\frac{\hbar^2}{2 m} \nabla^2 \right) \phi_\gamma ({\bf r})
\end{eqnarray}
for the kinetic energy, and for the pseudopotential
\begin{eqnarray}
C_{\delta \gamma} = \int d {\bf r} d {\bf r}' \;
\phi_\delta({\bf r}') V_{ps}({\bf r}, {\bf r}')
\phi_\gamma({\bf r}) \; ,
\end{eqnarray}
where $V_{ps}$ is in general a non-local pseudopotential operator.
Clearly, $C_{\gamma \delta}$ does not depend on $L_{\alpha \beta}$ for
either operator, so this term does not change as $L_{\alpha \beta}$ is
varied. It is thus easy to see that
\begin{eqnarray}
\frac{\partial E_c}{\partial L_{\alpha \beta}} =
\left[ 6(SLC + CLS)_{\beta \alpha} -
4(SLSLC + SLCLS + CLSLS)_{\beta \alpha} \right],
\label{eq:ecab}
\end{eqnarray}
where $C$ is the matrix representation of the corresponding operator
(kinetic energy or pseudopotential) in the basis of the support
functions.

For the Hartree and exchange-correlation contributions,
denoted by $E_v$, we have that
\begin{eqnarray}
\frac{\partial E_v}{\partial L_{\alpha \beta}} =
\int d{\bf r} \, \frac{\delta E_v}{\delta n({\bf r})}
\frac{\partial n({\bf r})}{\partial L_{\alpha \beta}} \; .
\label{eq:evab}
\end{eqnarray}
The electron density $n({\bf r})$
is simply $2 \rho({\bf r}, {\bf r})$, so that:
\begin{eqnarray}
\frac{\partial n({\bf r})}{\partial L_{\alpha \beta}} =
2 \sum_{\gamma \delta} \phi_\gamma({\bf r})
\frac{\partial}{\partial L_{\alpha \beta}}
\left[ 3(LSL)_{\gamma \delta} - 2 (LSLSL)_{\gamma \delta}
\right] \phi_\delta({\bf r}).
\end{eqnarray}
In the case of the Hartree contribution, we have
\begin{eqnarray}
\frac{\delta E_{\rm H}}{\delta n({\bf r})} = e^2
\int d {\bf r}' \frac{n({\bf r}')}{\mid {\bf r} - {\bf r}' \mid} = \Phi({\bf
r}),
\end{eqnarray}
where $\Phi({\bf r})$ is the Hartree potential, while in the case of the
exchange-correlation contribution we have
\begin{eqnarray}
\frac{\delta E_{\rm xc}}{\delta n({\bf r})} =
\frac{d}{d n} \left[ n \epsilon_{\rm xc}(n)\right] = \mu_{\rm xc}({\bf r}),
\end{eqnarray}
where $\mu_{\rm xc}$ is the exchange-correlation potential. If we take
$V({\bf r})$ to represent either $\Phi({\bf r})$ or
$\mu_{\rm xc}({\bf r})$ as the
case may be, we see that expression~(\ref{eq:evab}) reduces to
\begin{eqnarray}
\frac{\partial E_v}{\partial L_{\alpha \beta}} =
\left[ 6(SLV + VLS)_{\beta \alpha} -
4( SLSLV + SLVLS + VLSLS)_{\beta \alpha}\right],
\end{eqnarray}
where
\begin{eqnarray}
V_{\alpha \beta} = \int d {\bf r} \; \phi_\alpha({\bf r}) V({\bf r})
\phi_\beta({\bf r}).
\end{eqnarray}
By comparing this expression with Eq.~(\ref{eq:ecab}), it is easy to see
that the partial derivative of the total energy with respect to
$L_{\alpha \beta}$ can be written more compactly as
\begin{eqnarray}
\frac{\partial E_{\rm tot}}{\partial L_{\alpha \beta}} =
\left[ 6(SLH + HLS)_{\beta \alpha} -
4(SLSLH + SLHLS + HLSLS)_{\beta \alpha}\right],
\label{eq:etotalab}
\end{eqnarray}
where $H_{\alpha \beta}$ is the sum of the corresponding matrix elements
of the kinetic, pseudopotential, Hartree and exchange-correlation
operators, i.e. the matrix representation of the Kohn-Sham Hamiltonian
in the basis of the support functions. Recall that in practice, we do
not vary $E_{\rm tot}$
but rather vary $\Omega = E_{\rm tot} - \mu N$ with respect to $L_{\alpha
\beta}$. However, it is trivial to obtain $\partial \Omega/\partial
L_{\alpha \beta}$ from Eq.~(\ref{eq:etotalab}) by simply substituting
the matrix elements of $H$ by those of $H - \mu S$. Once this is done,
Eq.~(\ref{eq:etotalab}) corresponds to Eq.~(\ref{eq:partialab}).

\subsection{Functional derivative of $E_{\rm tot}$ with respect to
$\phi_\alpha$}

According to Eq.~(\ref{eq:separable}), the density matrix, and hence the
total energy, can be regarded as a function of the quantities $\phi_\alpha$
and $K_{\alpha \beta}$. When $\phi_\alpha$ is varied, $E_{\rm tot}$
therefore varies firstly because of its direct dependence on
$\phi_\alpha$, and secondly because of the implicit dependence
of the $K_{\alpha \beta}$ matrix on $\phi_\alpha$ through its
dependence on the overlap matrix elements $S_{\alpha \beta}$
(see Eq.~(\ref{eq:k_of_l})); we
call these two types of variation type 1 and type 2.

To see how variations of type 1 behave, consider first the kinetic and
pseudopotential energies. The type 1 variation of either of these is
given by:
\begin{eqnarray}
(\delta E_c)_1 = 2 \sum_{\gamma \delta} K_{\delta \gamma} \: \delta \!
\int d {\bf r} \; \phi_\gamma ({\bf r}) \hat{C} \phi_\delta({\bf r}) \; ,
\end{eqnarray}
where $\hat{C}$ represents the kinetic energy or the pseudopotential
operator. The variation of the integral gives
\begin{eqnarray}
(\delta E_c)_1 & = & 2 \sum_{\gamma \delta} K_{\delta \gamma}
\int d {\bf r} \; \left( \delta \phi_\gamma \hat{C} \phi_\delta +
\phi_\gamma \hat{C} \delta \phi_\delta \right) \nonumber \\
& = & 2 \sum_{\gamma \delta} K_{\delta \gamma}
\int d {\bf r} \; \left( \delta \phi_\gamma \hat{C} \phi_\delta +
\delta \phi_\delta \hat{C} \phi_\gamma \right).
\end{eqnarray}
The last equality follows from the fact that $\hat{C}$ is a Hermitian
operator.
The type 1 variation of $E_c$ can therefore be expressed as:
\begin{eqnarray}
\left( \frac{\delta E_c}{\delta \phi_\alpha({\bf r})} \right)_1 =
4  \sum_\beta K_{\alpha \beta} (\hat{C} \phi_\beta)({\bf r}),
\label{eq:conphi1}
\end{eqnarray}
where $( \hat{C} \phi_\beta ) ( {\bf r} )$
represents the action of the operator $\hat{C}$ on
$\phi_\beta$ evaluated at the point ${\bf r}$.

Now consider the type 2 variation of $E_c$ due to the variation
of the overlap matrix elements.
The variation of $S_{\alpha \beta}$ is
\begin{eqnarray}
\delta S_{\alpha \beta} = \int d {\bf r} \, (\phi_\alpha \delta \phi_\beta
+ \delta \phi_\alpha \phi_\beta) \; ,
\end{eqnarray}
and the type 2 variation of $E_c$ is then obtained by applying
this expression to Eq.~(\ref{eq:energycont}).
After a little manipulation, one obtains:
\begin{eqnarray}
\left( \frac{\delta E_c}{\delta \phi_\alpha ({\bf r})} \right)_2 =
12 \sum_\beta (LCL)_{\alpha \beta} \phi_\beta({\bf r}) -
8 \sum_\beta (LSLCL + LCLSL)_{\alpha \beta} \phi_\beta({\bf r}) \; .
\label{eq:conphi2}
\end{eqnarray}
Here, $C$ is the matrix whose elements are:
\begin{eqnarray}
C_{\alpha \beta} = \int d{\bf r} \, \phi_\alpha \hat{C} \phi_\beta \; .
\end{eqnarray}
Combining
Eqs.~(\ref{eq:conphi1}) and~(\ref{eq:conphi2}), we obtain the
following expression for the total variations of the kinetic and
pseudopotential energies:
\begin{eqnarray}
\frac{\delta E_c}{\delta \phi_\alpha ({\bf r})} =
4 \sum_\beta \left[ K_{\alpha \beta} \hat{C} + 3(LCL)_{\alpha \beta} -
2(LSLCL + LCLSL)_{\alpha \beta}\right] \phi_\beta({\bf r}).
\label{eq:de_dphi_kps}
\end{eqnarray}

For the remaining terms (Hartree and exchange-correlation), variation
in the energy results from variation in the electron density. Thus we
need to calculate
\begin{eqnarray}
\frac{\delta n({\bf r}')}{\delta \phi_\alpha({\bf r})} =
\frac{\delta}{\delta \phi_\alpha({\bf r})} 2 \sum_{\beta \gamma}
\phi_\beta({\bf r}') [3(LSL)_{\beta \gamma} -
2 (LSLSL)_{\beta \gamma}] \phi_\gamma({\bf r}').
\end{eqnarray}
Again, we will have variations coming directly from the change in
$\phi_\alpha({\bf r})$ and variations coming indirectly
from changes in the overlap matrix elements.
The total variation of $n({\bf r}')$ will be
\begin{eqnarray}
\frac{\delta n({\bf r}')}{\delta \phi_\alpha({\bf r})} &=& 2 \delta({\bf r} -
{\bf r}') \sum_\beta [\phi_\beta({\bf r}) K_{\beta \alpha} +
K_{\alpha \beta} \phi_\beta({\bf r})] \nonumber \\
& & + \; 2 \sum_{\beta \gamma}\phi_\beta({\bf r}')\phi_\gamma({\bf r}')
\frac{\delta}{\delta \phi_\alpha({\bf r})}
[3(LSL)_{\beta \gamma} - 2 (LSLSL)_{\beta \gamma}] \; .
\end{eqnarray}
Substituting this expression into
\begin{eqnarray}
\frac{\delta E_v}{\delta \phi_\alpha({\bf r})} =
\int d {\bf r}' V({\bf r}') \frac{\delta n({\bf r}')}{\delta \phi_\alpha({\bf
r})} \; ,
\end{eqnarray}
where the quantity
$V({\bf r}) \equiv \delta E_v / \delta n({\bf r}')$ represents the Hartree
or exchange-correlation potential, we find, after some manipulation:
\begin{eqnarray}
\frac{\delta E_v}{\delta \phi_\alpha({\bf r})} = 4 \sum_\beta
[K_{\alpha \beta} V ({\bf r}) +
3(LVL)_{\alpha \beta} - 2 (LSLVL + LVLSL)_{\alpha \beta}]
\phi_\beta({\bf r}) \; .
\end{eqnarray}
Combining this expression for the Hartree and exchange-correlation
derivatives with Eq.~(\ref{eq:de_dphi_kps})
for the kinetic and pseudopotential
derivatives, we find:
\begin{eqnarray}
\frac{\delta E_{\rm tot}}{\delta \phi_\alpha({\bf r})} = 4 \sum_\beta
[K_{\alpha \beta} \hat{H} +
3(LHL)_{\alpha \beta} - 2 (LSLHL + LHLSL)_{\alpha \beta}]
\phi_\beta({\bf r}) \; ,
\label{eq:de_dphi}
\end{eqnarray}
where $\hat{H}$ is the Kohn-Sham operator, and $H$ is its matrix
representation in the basis of support functions.

Note that in the practical grid-based calculations, the derivative we
actually want is $\partial E_{\rm tot} /
\partial \phi_\alpha ( {\bf r}_\ell )$,
which describes the variation of $E_{\rm tot}$ with respect
to change of $\phi_\alpha$ at the grid point ${\bf r}_\ell$. The formula
for $\partial E_{\rm tot} / \partial \phi_\alpha ( {\bf r}_\ell )$ is
identical to Eq.~(\ref{eq:de_dphi}) except that we need to multiply
by the volume per grid point $\delta \omega$.

\begin{figure}
\caption{(a) Total energy per atom as a function of the support
region radius $R_{reg}$ with $R_L$ = 5~\AA. (b) Total energy per
atom as a function of the range of the $L$ matrix, $R_L$, for two
different support region radii, $R_{reg} =$~2.04 and 2.38~\AA.}
\end{figure}

\begin{figure}
\caption{Support functions after minimisation of the total energy with
$R_{reg} = $~3.05 and $R_L =$~5.0~\AA. (a) $s$-like support
function, and (b) $p_x$-like support function.}
\end{figure}

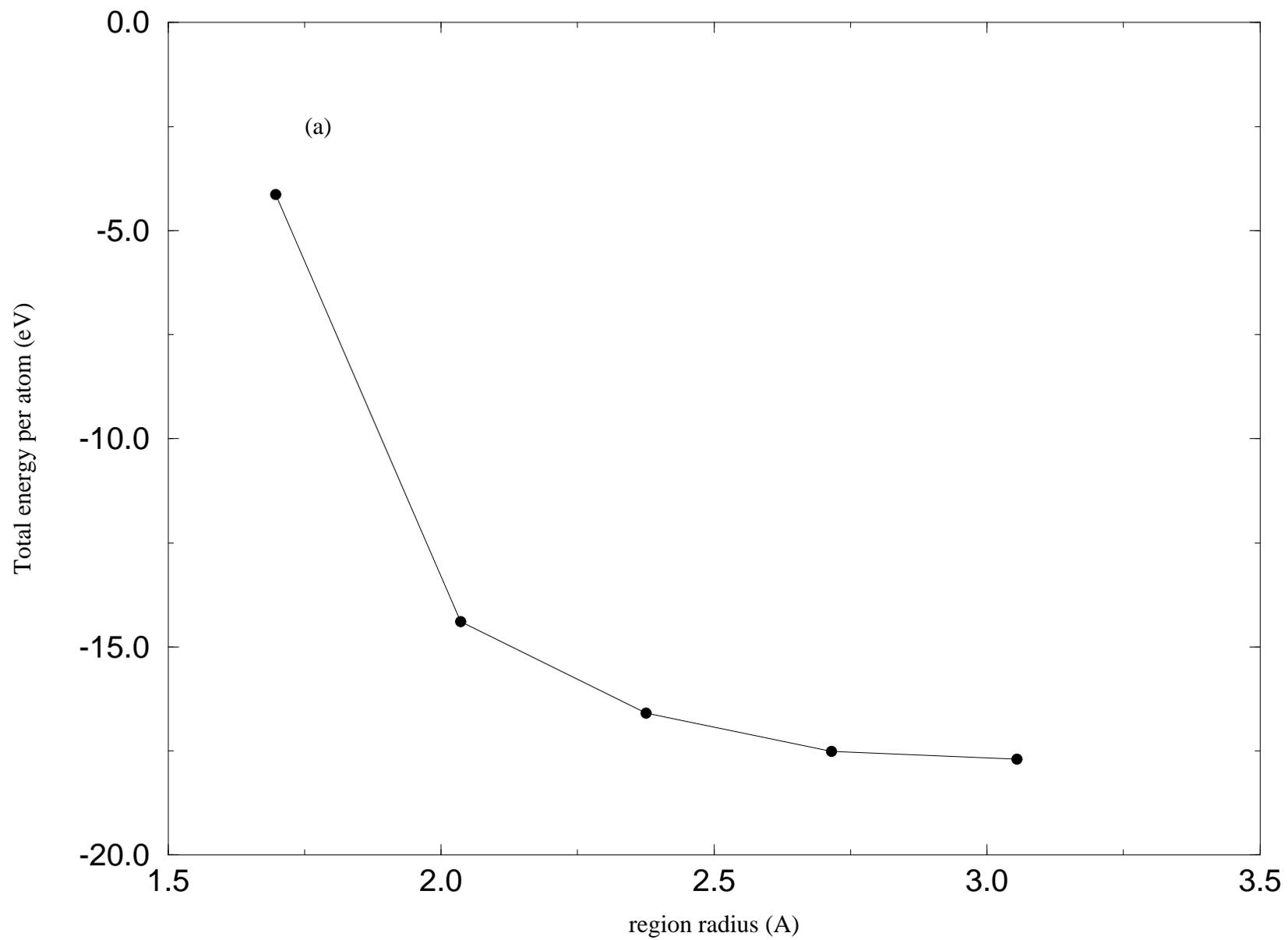
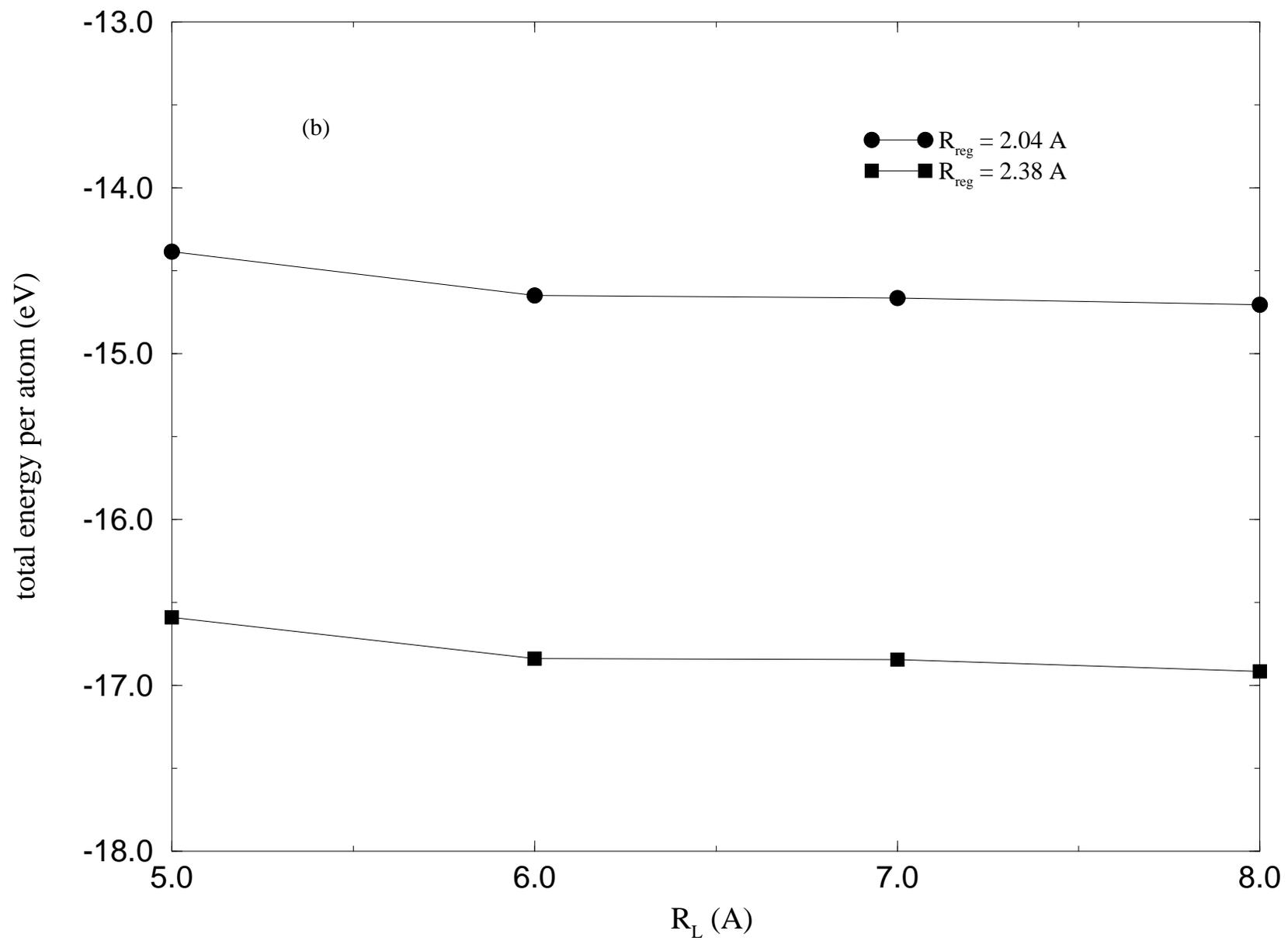
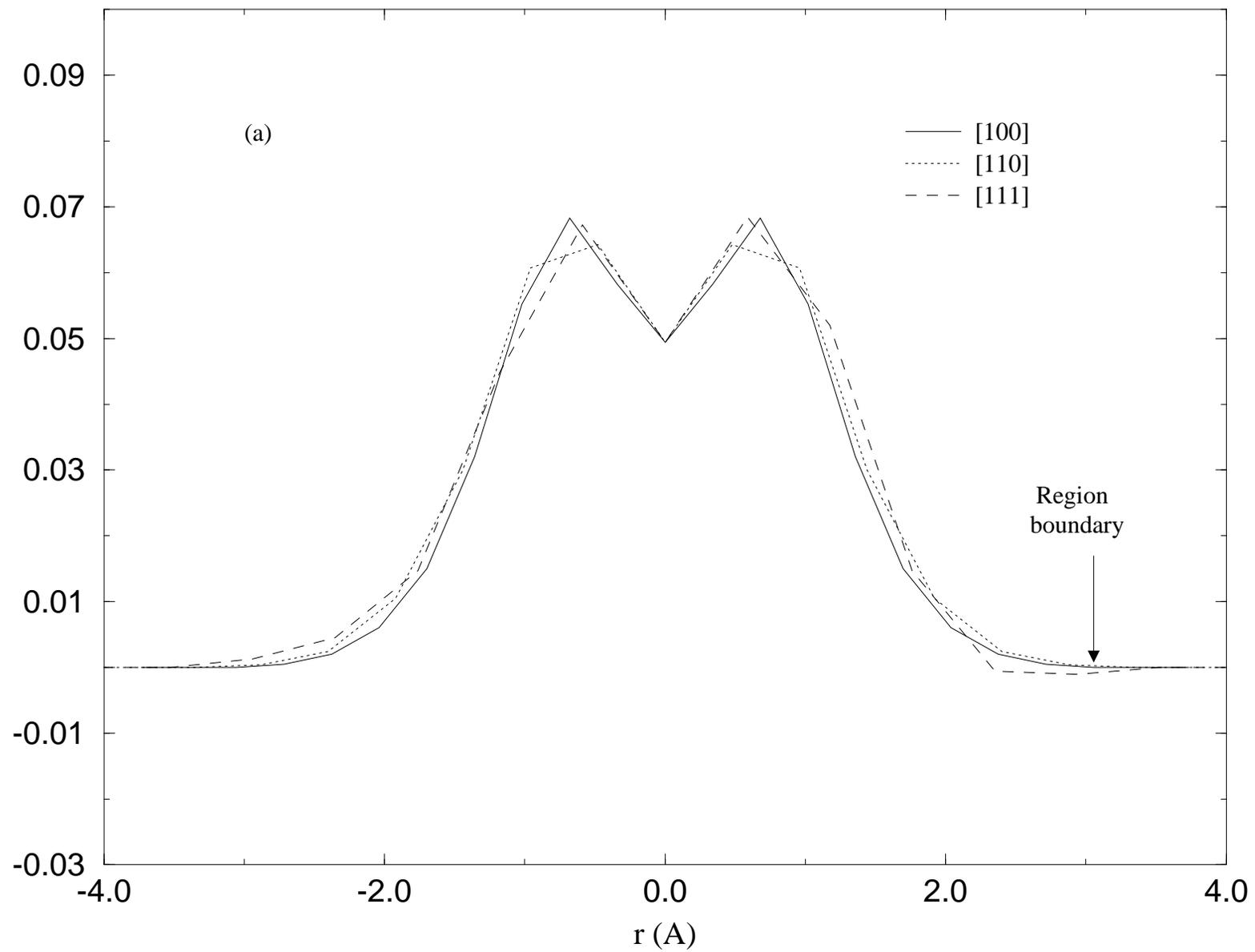
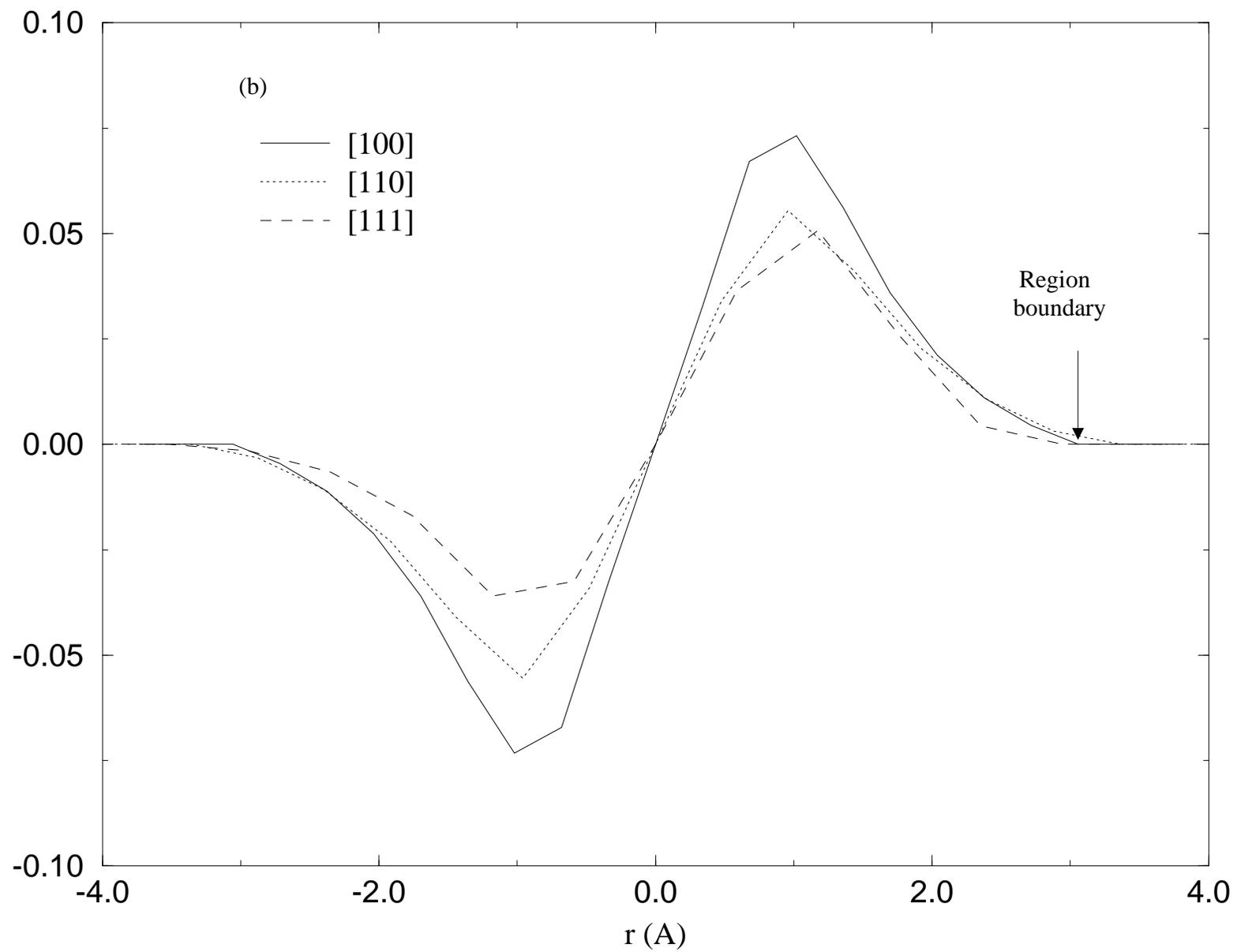
\end{document}